\documentclass[review]{elsarticle}

\usepackage{hyperref}
\usepackage{color}

\usepackage{listings}
\definecolor{mygreen}{RGB}{28,172,0} 
\definecolor{mylilas}{RGB}{170,55,241}
\journal{Ecological Indicators}

\begin{document}

\begin{frontmatter}

\title{Extended patchy ecosystems may increase their total biomass through self-replication}

\author{Mustapha Tlidi\fnref{emialMT}}
\address{Facult\'e des Sciences, Universit\'e Libre de Bruxelles (ULB), Campus Plaine, CP. 231, B-1050 Bruxelles, Belgium.}
\fntext[emialMT]{mtlidi@ulb.ac.be}

\author{Ignacio Bordeu\fnref{emailIB}}
\address{Department of Mathematics and Centre for Doctoral Training on Theory and Simulation of Materials, Imperial College London, 180 Queen’s Gate, London SW7 2AZ, United Kingdom.}
\fntext[emailIB]{ibordeu@imperial.ac.uk}
\author{Marcel G. Clerc}
\address{Departamento de F\'isica, Facultad de Ciencias F\'isicas y Matem\'aticas,
Universidad de Chile, Casilla 487-3, Santiago, Chile.}

\author{ Daniel Escaff}
\address{Complex Systems Group, Facultad de Ingenier{\'i}a y Ciencias
Aplicadas, Universidad de los Andes, 
Monse{\~n}or Alvaro del Portillo  12.455, Las Condes, Santiago, Chile.}

\begin{abstract}
Patches of vegetation consist of dense clusters of shrubs,  grass, or trees, often found to be circular characteristic size, defined by the properties of the vegetation and terrain. Therefore,  vegetation patches can be interpreted as localized structures. Previous findings have shown that such localized structures can self-replicate in a binary fashion, where a single vegetation patch elongates and divides into two new patches, in a process  resembling cellular mitosis. Here, we extend these previous results by considering the more general case, where the plants interact non-locally, this extension adds an extra level of complexity and shrinks the gap between the model and real ecosystems, where it is known that the plant-to-plant competition through roots and above-ground facilitating interactions have non-local effects, i.e. they extend further away than the nearest neighbor distance. Through numerical simulations, we show that for a moderate level of aridity, a 
transition from a single patch to periodic pattern occurs. Moreover, for large values of the hydric stress, we predict an opposing route to the formation of periodic patterns, where a homogeneous cover of vegetation may decay to spot-like patterns. The evolution of the biomass of vegetation patches can be used as an indicator of the state of an ecosystem, this allows to distinguish if a system is in a self-replicating or decaying dynamics.
In an attempt to relate the theoretical predictions to real ecosystems, we analyze landscapes in Zambia and Mozambique, where vegetation forms patches of tens of meters in diameter. We show that the properties of the patches together with their spatial distributions are consistent with the self-organization hypothesis. We argue that the characteristics of the observed landscapes may be a consequence of patch self-replication, however, detailed field and temporal data is fundamental to assess the real state of the ecosystems.
\end{abstract}

\end{frontmatter}

\section{Introduction}
Spontaneous shift from a uniform cover of vegetation into a fragmented ecosystem 
constituted by a spatially periodic distribution of gaps, or patches is an irreversible process. 
This transition may occur either in  water or nutrient limited territories. 
It is now  widely admitted that  facilitative and competitive interactions 
between individual plants can directly or indirectly account for the formation of vegetation patterns  \cite{Lefever1997,LT,Lejune_couteron,Klausmeier1826,Rietkerk_2001,Meron2001,QUA:QUA10878,sherratt2005analysis,d2006patterns,vesipa2015noise}. The spatial distribution of such patterns can be modified the effect of climatic, terrain and anthropogenic influences \cite{Kefi_07,Dakos_11,Deblaw_12}. Quantitative studies based on field observation have been made on: the Sahelian  gapped patterns, constituted by Combretum micranthum trees  \cite{JEC:JEC1126,ECY:ECY20088961521,Lefever2009194}; patches of vegetation in arid high altitude environments in the tropical
alpine ecosystems of the Andes formed by  Festuca orthophylla (Poaceae); and for grasses or by Pycnophyllum tetrastichum cushions (Bolivia, \cite{Couteron20140102}). 

\begin{figure}[ht!]
\centering
\includegraphics[width=12 cm]{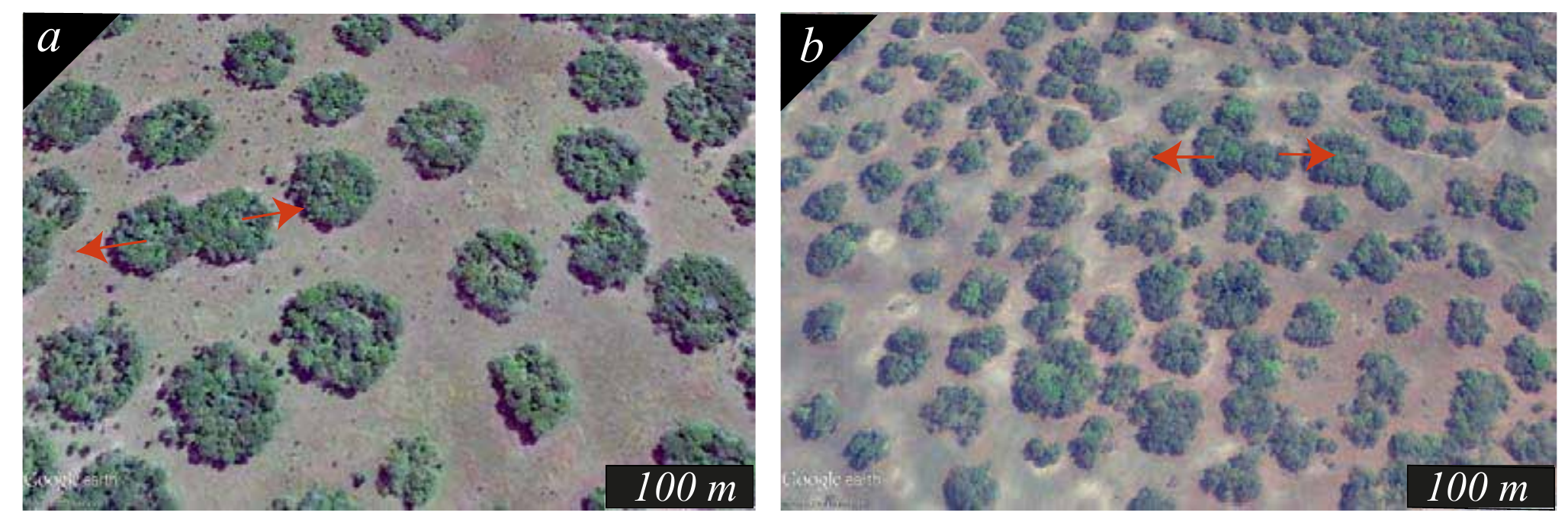}
\caption{(color online) {\bf Satellite images} ({\it {Google Earth Pro}}), of vegetation patches in a) 
the  Mufumbwe District in the North-Western Province of Zambia [$ 13^{\circ} 46'39.83''$S, $ 25^\circ16'39.59''$E],  and b) the Fombeni, Mozambique [$18^\circ41'02.17''$S, $35^\circ 31'55.95''$E]. 
The red arrows indicate overlapping patches, possibly undergoing self-replications}.
\label{Figure-1}
\end{figure}
Patterned vegetation landscapes are fragmented, i.e, the terrain is only partially covered by vegetation. Some of these landscapes are composed of vegetation patches, which may be sparsely or regularly distributed. These patches usually have a characteristic size and well defined circular shape.
It has been shown recently that localized patches can be destabilized by a deformation of their circular
shape, either leading to the formation of labyrinthine patterns \cite{bordeu2016laby}, or  dividing into two 
new identical patches  of smaller diameter \cite{bordeu2016self}. 
The latter is a phenomenon often called self-replication and resembles mitotic cell division. It has been studied in the context of herbaceous populations in arid ecosystems \cite{bordeu2016self}. The self-replication mechanism allows the transition from a single localized structure into a qualitatively different state, namely a hexagonal periodic pattern of vegetation. During the transition from localized to periodic pattern, the total biomass increases as newly formed patches contribute to the repopulation of the territory accessible to vegetation. From a theoretical point of view, self-replication is a patterning phenomenon better known in  physico-chemical contexts rather than ecological systems. 
It is a generic mechanism of pattern formation, which has been observed and established in various non-equilibrium systems, such as fluids \cite{Magnetic}, liquid crystals \cite{refId0,Oswald200067}
chemical systems \cite{pearson:93,lee1994experimental,PhysRevLett.79.1941,kaminaga2005black,PhysRevLett.98.188303,DeKepper,monine2002modeling,Schaak1998386,PhysRevLett.81.1726,Tlidi_Gandica_16}, in plant ecology \cite{meron2004vegetation,bordeu2016self}, 
material sciences \cite{Ren_2003,Yasumasa}, granular fluid systems \cite{PhysRevLett.99.038001,sandnes2011patterns} and nonlinear optics \cite{PhysRevLett.89.233901}.

In this contribution, we investigate the space-time dynamics of vegetation under a self-replication phenomenon by extending the previous work by Bordeu et al. \cite{bordeu2016self,bordeu2016laby}, where a simple local model was used to illustrate that self-replication was a possible mechanism for vegetation propagation. Here we consider general integro-differential model instead of the simplified model of \cite{bordeu2016self,bordeu2016laby}, here, non-local interactions are taken explicitly  into account. The non-localities arise from the facilitating mechanisms, i.e. promoters of vegetation growth, competition mechanisms, which limit vegetation growth, and dispersion effects. This model corresponds to a variant of the theory of vegetation patterning
established by R. Lefever \cite{Lefever1997}, which focuses on the relationship between the structure 
of individual plants and the facilitation-competition interactions existing within plant communities. 
It is now widely recognized that the existence of facilitation and competition interactions play an important role 
in the formation of self-organized vegetation patterns. Numerical simulations of our  model show indeed a self-replication process  that leads moderately arid ecosystems to undergo a transition to higher biomass states, namely hexagonal patterns of vegetation patches. Moreover, we show that this kind of patterns may be obtained through the decay of a homogeneous vegetated landscape towards a less populated  fragmented state, where hydric stress induces  contraction of vegetated areas.
Depending on the levels of aridity, the ecosystems may decay to a different type of patterned states or even become desert.

We study the characteristics of both self-replication and fragmentation processes through the analytic and numerical analysis of a general integro-dfferential model. We show, from a theoretical perspective, that depending on the levels of aridity 
localized patches can be more or less stable than the periodic pattern, a phenomenon previously studied in simpler local models \cite{PhysRevA.84.043848}. In an attempt to conciliate the theoretical observations with real data, we consider two ecosystems, namely, from Zambia and Mozambique, these landscapes are composed of vegetation patches reaching large sizes, of the order of tens of meters in diameter (see Fig.~\ref{Figure-1}). We perform statistical 
analysis of satellite images and find that patches have correlated characteristic patch sizes and inter-patch distances along with other properties that support the hypothesis of the self-organization nature of these landscapes. 

The article is organized as follows: in Section \ref{Sec-Model} the theoretical model is introduced, together with  the phase diagram. A description of the methods used to analyze the satellite images are included in section \ref{Sec-Methods}. The results are presented in section \ref{Sec-Results}. Theoretical results indicating the relationship with the wavelength and the range of the facilitative and competitive interactions are presented in the appendix. Finally, we present the Discussion, Conclusions and Perspectives of our work. 

\section{Methods}

\subsection{Mathematical  model} \label{Sec-Model}
The modeling of ecosystems is a challenging and complex problem. Here, we adopt the
theory of vegetation patterns established by R. Lefever and coworkers two-decades ago to model the spatiotemporal 
dynamics of vegetation in which both space and in time are considered to be continuous variables 
\cite{Lefever1997,LT,Lejune_couteron}. This theory incorporates the non-local facilitative and the competitive plant-to-plant interactions though kernels \cite{Lefever1997,Tlidi2008,Lefever2009194,Lefever-Turner}. In the absence of these interactions, 
the resulting model is similar to  the paradigmatic logistic equation introduced by  Verhulst to study  
population dynamics \cite{Verhulst1845,mawhin2002heritiers,mawhin2004legacy}. In what follows we  consider vegetation of a single specie  
settled on a flat landscape under isotropic and homogeneous environmental conditions. To simplify further 
the description of the system, we assume  that  all plants are mature. Thus, we neglect age 
classes. This approximation can be justified by the fact that individual plants grow on much faster time scale 
comparing to the time scale of the formation of regular vegetation pattern. The only  variable is the 
vegetation biomass density which is defined at the plant level.  Let us introduce  the biomass density, $b({\bf{r}},t)$,
 that satisfies the following dynamical evolution 
\cite{Tlidi2008,Lefever2009194}
\begin{equation} \label{nonlocal}
 \partial_t b({\bf{r}},t) = b({\bf{r}},t)\left[1-b({\bf{r}},t) \right]M_f({\bf{r}},t) -\mu b({\bf{r}},t)M_c({\bf{r}},t)+DM_d({\bf{r}},t),
\end{equation}
where $\bf{r}$ and $t$ are the spatial coordinates and time, respectively. The time derivative is represented by $\partial_t$. 
The parameter $\mu$, is the decay-to-growth~rate ratio. It can be viewed as an indirect measure of resource scarcity or stress, that limits net biomass production
and is what we refer to as aridity parameter. The first and the second terms on the right-hand-side of Eq. (1) 
account for the plant-to-plant facilitation and competition feedbacks, respectively. 
They describe the spatial extension of feedback effects in terms of the characteristic ranges $L_f$ and $L_c$ 
over which facilitative and competitive interactions operate, respectively. 
The facilitative interaction acts on the level of the aerial plant structure (crown) 
that involves sheltering, litter, water funneling or any other effect, such as seed production 
and germination that contribute to the biomass natural growth \cite{JEC:JEC1295}.
Let $L_a$ be the crown radius projected on the 
surface  element $A=\pi L_a$ centered on a point $\bf{r}$. Assuming  that the length of facilitative 
interaction is equals to the radius of the crown, i.e.,  $L_f=L_a$. On the other hand, in resource-limited 
environments, plants should compete for their survival. In face of  climate change and increasing drought 
periods, plants should adapt their root structures to overcome resource scarcity \cite{JEC:JEC682,JEC:JEC1126}. 
Through their root structures (rhizosphere), each individual plant tends to  deprive its neighbors of 
vital resources, such as water or nutrient uptake. Measurements of roots lateral spread indicate 
that they extend beyond the radius of the aerial structure (crown) by an order of magnitude \cite{JEC:JEC1126,ECY:ECY20088961521,Lefever2009194,Couteron20140102}.

The competitive interaction between plants tends to oppose the facilitation mechanisms, 
by impeding vegetation growth. In arid landscapes, the length of facilitative interaction 
(crown) is much shorter than the length of the competitive interaction operating at the 
level of the rhizosphere which is the volume of soil around living roots. Let consider 
also that the range of the competition is such that  $L_c=L_r$, where, $L_r$ is the 
radius of the rhizosphere. The third term in Eq. (1) describes  the spatial 
propagation of vegetation via seed dispersion. The parameter $D$ is the rate of propagation of the vegetation. 
The competitive plant-to-plan interaction is considered to be of the form
\begin{equation}
\label{Ker_fac}
M_{c}({\bf{r}},t) =\exp \biggl(\frac{\chi_{c}}{N_c}\int e^{-|{\bf{r^{\prime}}}|/L_c}\,b(|{\bf{r}}+{\bf{r}^\prime}|,\, t)\,d\mathbf{r^{\prime}} \biggr),
 \end{equation} 
where, $\chi_c$ is the strength of the competitive  interaction, and  $N_c$ is a normalization constant, that depends on the spatial
dimension. In two dimensions, $N_c=2\pi L_c^2$. The spatial propagation of vegetation via seed dispersion is assumed to have the form
\begin{equation}\label{Ker_dis}
M_{d}({\bf{r}},t) =\frac{\delta}{\pi}\int e^{-|{\bf{r^{\prime}}}|^2/L_d^2}\, [b({\bf{r}}+{\bf{r}^\prime},\, t)-b({\bf{r}}\, t)]\,d\mathbf{r^{\prime}},
\end{equation}     
the parameter $L_d$ and $\delta$ are, respectively, the dispersion range of seeds  
and the strength of dispersive process.  To simplify further the analysis we assume that 
the seed dispersion is described as diffusion $M_{d}({\bf{r}},t) \approx D\nabla^2b({\bf{r}},t)$. This can be obtained by 
 considering a small dispersion range and a simple Taylor expansion, which leads to $D = \delta L_d^4 /4$. 
To simplify further the problem, we consider that the facilitation is well described by a local process modelled 
by $M_{f}({\bf{r}},t) =\exp(\chi_{f}b)$, where $\chi_f$ is the strength of the facilitative interaction.

The analysis we make in this work generalizes previous results. Here, the complete non-local integro-differential model is analyzed, which accounts for an important step forward in the understanding of the behavior of this type of system. For a complete linear stability analysis of this model see Appendix~\ref{An-1}.

\subsection{Localized vegetation patches}
\begin{figure}[htbp!]
\centering
\includegraphics[width=12 cm]{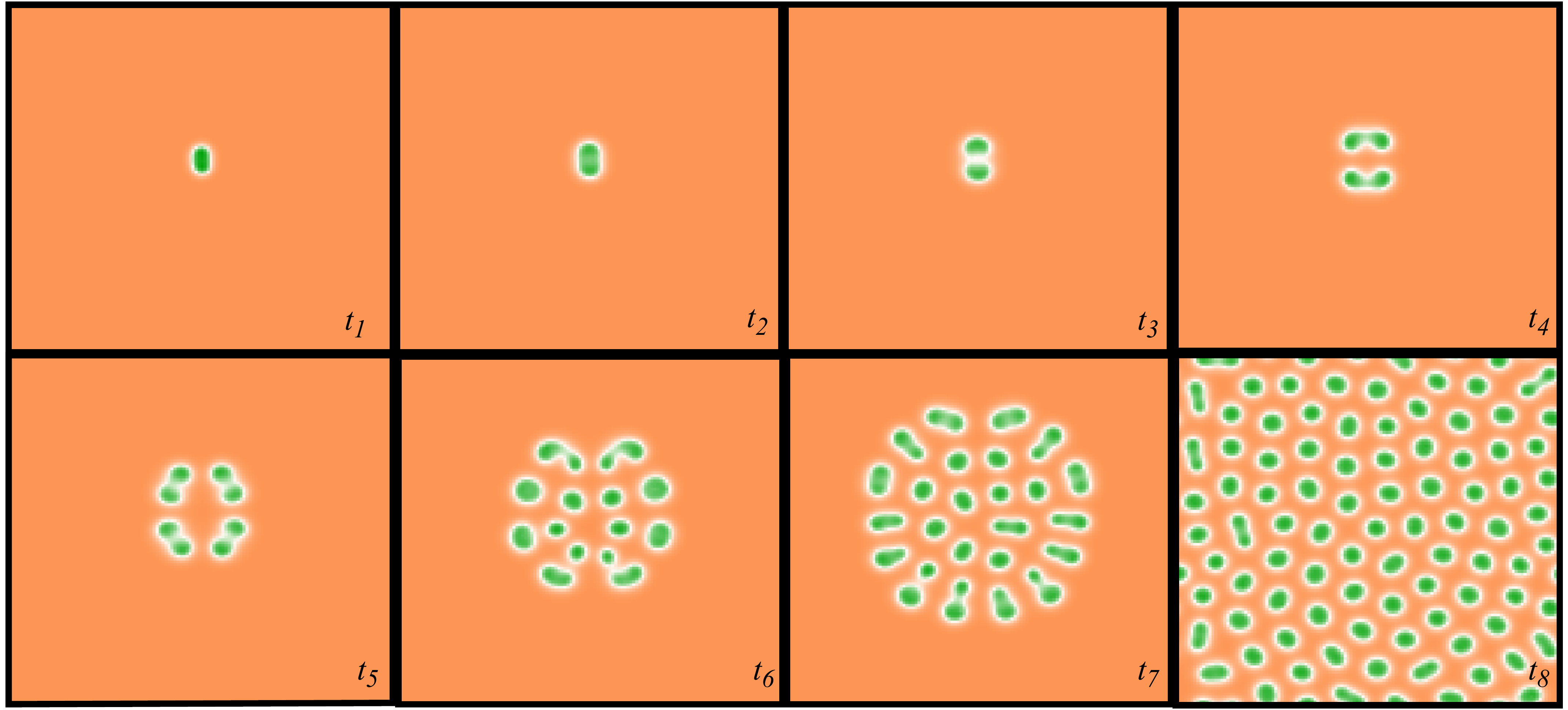}
\caption{(color online)  {\bf Evolution of a self-replicating localized vegetation patch}, for a sequence of time points   
$t_1<t_2<...<t_5<t_6$. Obtained through direct simulation of Eq.~\ref{nonlocal}, for a system of size $128\times128$-points, and parameters: $\mu=1.02$, $\chi_f=2$, $\chi_c=1$, $L_c=4.5$, $D=1$, $dt=0.1$,  and $dx = 1.0$.}
\label{Figure-4}
\end{figure}

The non-local equation~(\ref{nonlocal}) exhibits stable circular localized structures which
are supported by one of the homogeneous steady state, $b_0=0$ (unpopulated state) or $b_s$ (homogeneously vegetated). In the context of 
vegetation dynamics, localized structures that emerge from the unpopulated 
state $b_0=0$, correspond to circular patches of vegetation which are surrounded 
by bare terrain. As previously mentioned, periodic states and localized {patches emerge as a self-organizing response of the system to changes in the parameters. Aridity ($\mu$), is of particular interest, as it can be directly related to field measurements, and also relates to an ever-increasing concern due to global warming. An increase in aridity (lower hydric resources available or other stress factors) may cause a contraction of savannas and woodlands, in a process called desertification, triggering the formation of vegetation patterns or even deserts. 
The necessary and sufficient condition
for the formation of localized patches is the coexistence between a homogeneous cover and a periodic vegetation pattern. Implying the existence of an hysteresis loop. Inside this loop, there is a so-called pinning range of the aridity parameter where localized gaps or patches are stable \cite{Lejeune-Tlidi-Couteron,rietkerk2004self,Tlidi2008}. Similar pinning behavior occurs in many spatially extended systems where
a homogeneous steady state coexists with a spatially periodic state \cite{POMEAU19863,PhysRevLett.73.640,PhysRevLett.84.3069}.
Pinning was first reported for front solutions by Y. Pomeau \cite{POMEAU19863}, has also been applied to gap vegetation patterns \cite{PhysRevA.84.043848}.
Localized structures and localized patterns are a well 
documented phenomenon, concerning almost all fields of natural science including chemistry, biology, ecology, physics, 
fluid mechanics, and optics \cite{Tlidi_Focus_chaos_07,DSbook,Rev18,Tlidi20140101,knobloch2015spatial,Meron_book,tlidi2015nonlinear}.  
It is worth mentioning that fairy circles are striking examples attributed to this category of 
localized vegetation patterns \cite{vanroyen}. Although, the mechanisms leading to their formation are still subject of debate among the scientific community. 
They have been interpreted as the result of a pinning front between a uniform plant distribution 
 and a periodic hexagonal vegetation pattern  \cite{Tlidi2008}. Recent investigations support this interpretation \cite{tschinkel2012life,cramer2013namibian,tschinkel2012life,getzin2015adopting}. 
On the other hand, fairy circles formation may result from front dynamics connecting a bare state and uniform plant distributions \cite{fernandez2014strong,fernandez2013strong,PhysRevE.91.022924}. In all these works, the origin of fairy circle formation is intrinsic to the dynamics of the system. This means that the diameter of the fairy circle is  determined rather by the parameters of the system and not by external effects, such as the presence of social insect or anisotropy. However,  another theory based on external effects, such as termites or ants has been suggested \cite{Becker,Picker,Juergen,Bonachela}. More recently, Tarnita  and collaborators have shown that a combination of intrinsic and extrinsic effects could explain the origin of fairy circles \cite{Tarnita}.

In addition to analytically deriving predictions from Eq.~\ref{nonlocal}, we will provide a case study for which the interpretation of vegetation patchiness based on localized structures and self-replication is plausible.
\subsection{Data analysis}
\label{Sec-Methods} 
Through the use of satellite images obtained from {\it Google Earth}, we have located regions in Zambia and Mozambique where vegetation patches dominate the landscape (see SI Table~\ref{Table-PP} for details).

\subsection*{Patch detection algorithm}
In order to detect the vegetation, we developed a simple algorithm that segments the images, detects vegetation patches and extract properties of interest (Matlab vR2016b, see Supplementary Information \ref{Code}). The boundaries of every detected patch are used to extract the patch geometrical features, such as area, perimeter, equivalent diameter, and centroid positions, which are then used in the spatial analysis.
\subsection*{Equivalent radius and nearest-neighbor distance}
The equivalent radius of each structure $r_{eq}$ is calculated as
\begin{equation}
\label{Eq-Req}
r_{eq} = \sqrt{\frac{A_s}{\pi}}
\end{equation}
where $A_s$ corresponds to the area of the structure. The nearest neighbor distances is obtained by finding the minimum of the distance every patch and all the other patches.

\subsection*{Spatial distribution analysis}
\subsubsection*{L-function}
For the analysis of the spatial distribution of patches we make use of the modified Ripley's function $L$, for this, only the centroids of the patches are used. We define $L$ as \cite{Ripley_76} 
\begin{equation}
\label{Eq-L}
L(r) = \sqrt{\frac{A}{\pi N^2} S(r)}
\end{equation}
where $r$ is the distance from the reference point. $N$ and $A$ are the total number of points and the area, respectively. $S(r)$ is the number of points that lay inside a circle of radius $r$ centered on a patch of reference chosen randomly, this is repeated for multiple patches.
As control we compute the $L$ function for a set of randomly distributed point (null model), by repeating this 200 times, construct the $95\%$ confidence intervals. When the observed values of L go above (below) the confidence interval, we say the distribution of point is clustered (dispersed).
\subsubsection*{Radial distribution function}
Similarly to the L-function, the radial distribution function $g(r)$, is computed using only the positions of the centroids of the detected objects, we define $g(r)$ as \cite{Couteron_Kokou_97}
\begin{equation}
\label{Eq-g}
g(r) = \frac{N(r)}{2\pi r \Delta r \rho}
\end{equation}
where $N(r)$ is the number of points that lie inside a ring of radius $r$ and width $\Delta r$, centered on a given object. The numerator 
of this expression corresponds to the total number of points that lay inside the ring for the case of homogeneously distributed objects, $\rho$ is the overall number density. The radial distribution function, $g(r)$, measures the correlation of the position of the points, values of $g(r)>1$ ($g(r)<1$) indicate positive (negative) correlation. N(r) in numerically computed by measuring the distance between every two patches in the images, and then performing a histogram, where the bin-width corresponds to $\Delta r$. 
\subsubsection*{Fourier transform}
The Fourier transform is used considering the binary image $U(x,y)$ of the structures detected. It is defined as
\begin{equation}
\label{Eq-FT}
\hat{U}(k_x,k_y) = \int\int_{-\infty}^\infty U(x,y) \exp(-2\pi i (k_x x + k_y y)) dxdy
\end{equation}    
Where $(x,y)$ are the spatial coordinates of the image pixels, and $(k_x,k_y)$ are the spatial frequencies, usually $\hat{U}$ is referred to as the spectrum of $U$. Peaks in the spectrum indicate dominant wavelengths or typical sizes of the patches detected in the satellite images (see \cite{LT,Lejune_couteron,Debloaw_Couteron_2011}.

\section{Results} \label{Sec-Results}
  
\subsection{Self-replication: From vegetation patches to extended pattern}
Detailed mathematical analysis is provided for the first time for Eq.~\ref{nonlocal} (see the Appendix~\ref{An-1}), this analysis} shows that under a wide range of the aridity parameter, 
there is coexistence between homogeneous and pattern states. Thus, supporting stable localized vegetation patches. Moreover, as observed in models with only local interactions \cite{bordeu2016self}, vegetation patches in the non-local model may also be affected by a  Turing-Prigogine instability, where the patch elongates, increasing in size to finally split into two patches, through the decay of the central "bridge" that connected them. This {\it mitotic} dynamics occurs 
repeatedly to each of the new patches, allowing the single initial patch to end up covering the whole 
system, through the formation of a periodic hexagonal pattern, as can be observed in Fig.~\ref{Figure-4}.
It is through the self-replication mechanism that a single vegetation patch can induce a increase the total biomass of a system although the aridity is too high to allow for an homogeneous cover of vegetation, Figure~6 shows the increase of the total biomass on the transition from a single to four patches through self replication.

To study the parameter zone where self-replication is observed, we have performed a direct numerical approach 
for computing the rate at which the elongating unstable mode grows, see Fig.~\ref{Figure-5}. We have found that given a size-scale, $L_c$, for the nonlocal interactions in Eq.~(\ref{nonlocal}), there is a wide range of aridity, $\mu$, for which an initial localized structure will destabilize through the self-replicating mechanism (see Fig.~\ref{Figure-5}). Lower values of $L_c$, imply shorter-range nonlocal effects, this generates a left-shift in the self-replicating window to lower values of aridity as can be seen in Fig.~\ref{Figure-5} when comparing the red ($L_c=4.25$, $*$) and yellow ($L_c=4.00$, $\circ$) curves to the blue curve ($L_c=4.50$, $\times$). 

\begin{figure}[ht!]
\centering
\includegraphics[width=12 cm]{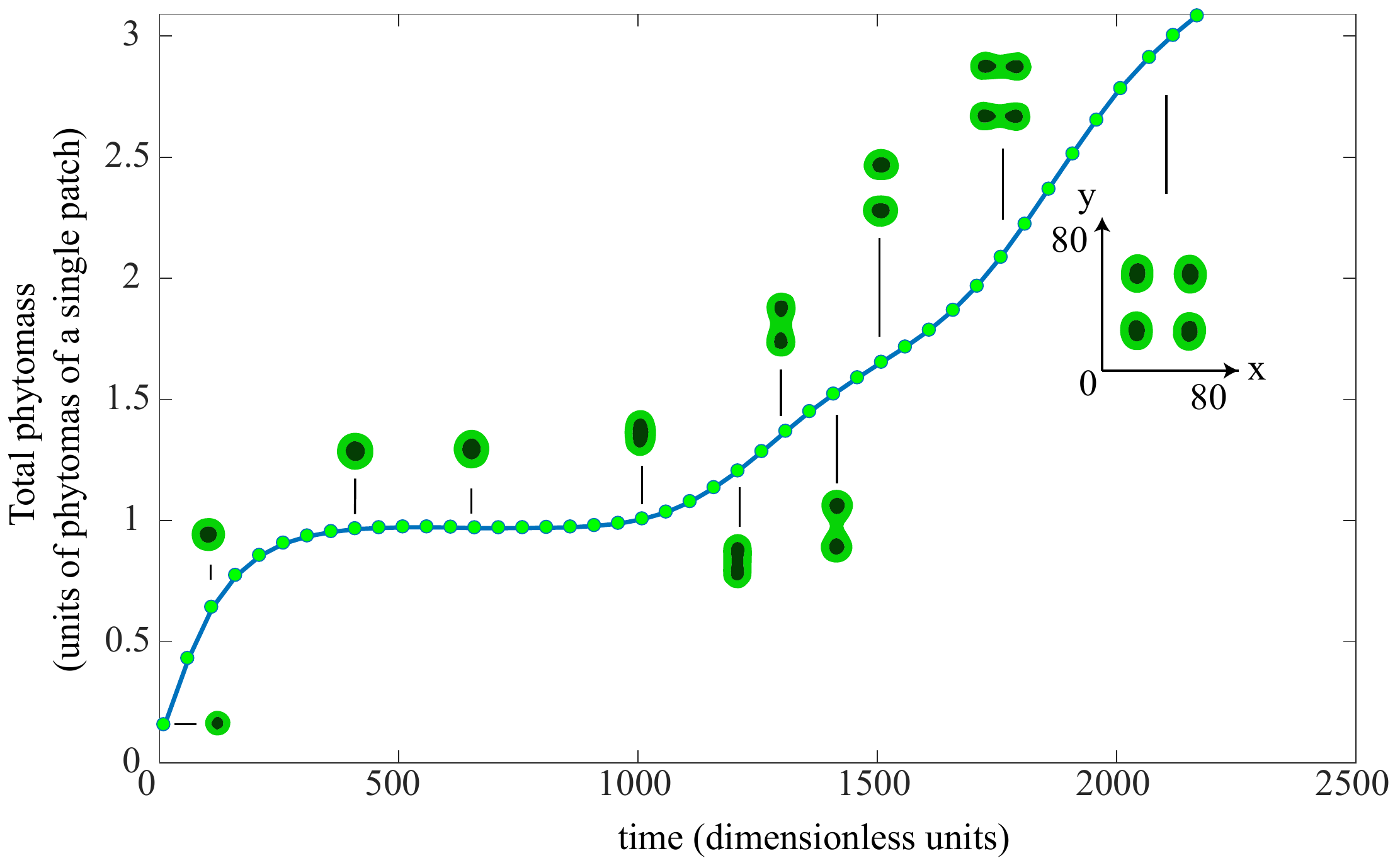}
\caption{(color online) {\bf Evolution of the total biomass} on the transition from a single to four patches through self replication. Points obtained through direct simulation of Eq.~\ref{nonlocal}, for a system of size $128\times128$-points, and parameters: $\mu = 1.02$, $L_c=4.50$, $\chi_f=2$, $\chi_c=1$, $D=1$, $dt = 0.1$, and $dx = 1.0$. Multiplicative noise amplitude $0.1$, facilitates the initial break of symmetry. The insets show the different stages of the evolution. The total biomass is normalized with respect to the total biomass of a mature patch. Thus, during maturation a single spot reaches a normalized biomass of 1, to replicate into two patches to reach a total of almost 2. After a second replication (four patches) the system reaches a normalized biomass of 3.}
\label{Evo1}
\end{figure}

Moreover, we have found that for aridity parameter values below the self-replicating region, 
there is a different route for the self-replicating mechanism (see ring instability inset, Fig.~\ref{Figure-5}), 
here, an initial localized structure grows radially. After reaching a critical radius, the central portion 
of the structure decays, forming a doughnut-like shape. This structure is also unstable. By the 
consequent decay of two opposite sides 
of the doughnut, the structure ends up dividing into two new localized patches. 

\begin{figure}[ht!]
\centering
\includegraphics[width=12 cm]{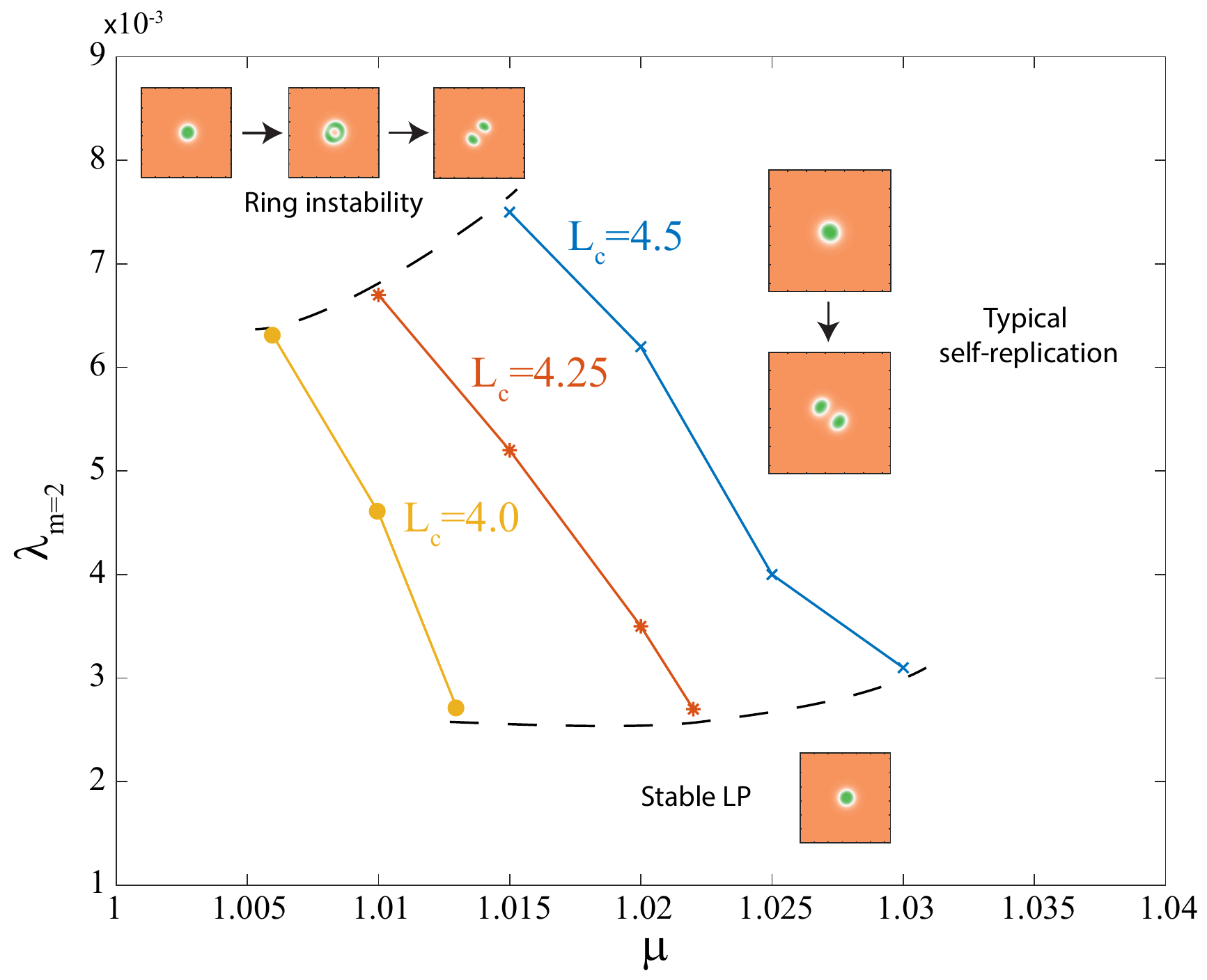}
\caption{(color online) {\bf Growth rate of the unstable mode}, corresponding to an elongation, leading to division. For: ($\times$) $L_c=4.50$; ($*$) $L_c=4.25$; ($\circ$) $L_c=4.00$, as a function of the aridity parameter $\mu$. Lines are linear interpolations. Outside the limits denoted by dashed lines, initial localized structures may remain stable (large $\mu$ values), or destabilize through a ring instability, leading to division (smaller $\mu$ values). Points obtained through direct simulation of Eq.~\ref{nonlocal}, for a system of size $128\times128$-points, and parameters: $\chi_f=2$, $\chi_c=1$, $D=1$, $dt = 0.1$, and $dx = 1.0$.  Multiplicative noise amplitude $0.1$, facilitate the initial break of symmetry.}
\label{Figure-5}
\end{figure}

\begin{figure}[ht!]
\centering
\includegraphics[width=8 cm]{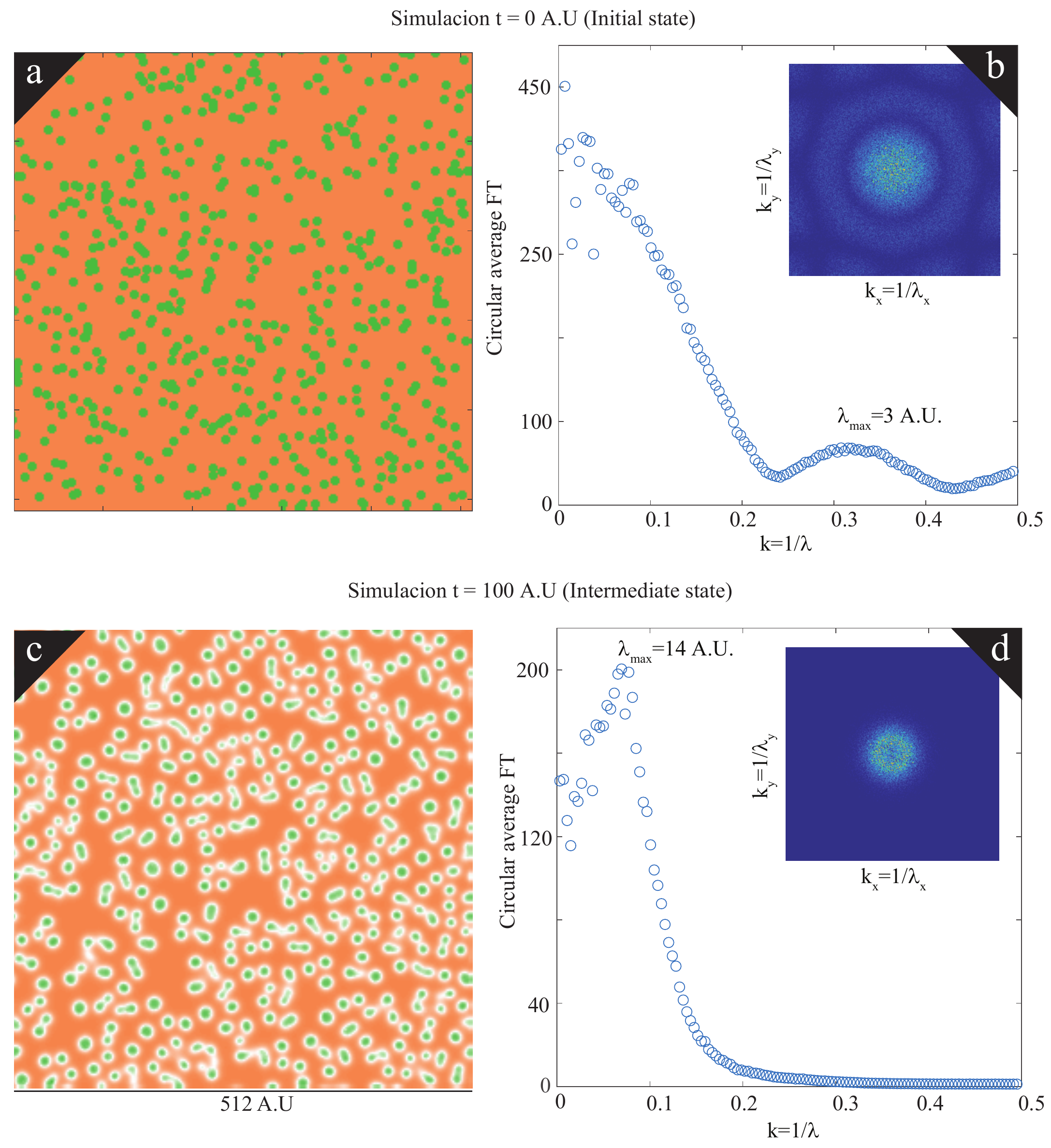}
\caption{(color online) {\bf Large scale simulations}, (a,b)-shows the $256\times256$-point ($dx=2.0$) initial condition generated from a Poisson point process with structures of $3$-point radius, and its corresponding Fourier transform, respectively. (c,d)-Shows the evolved state after 1000 iterations ($dt = 0.1$), and the corresponding Fourier transform, respectively. Parameter: $\mu=1.02$, $L_c=4.5$, $\chi_f=2$, $\chi_c=1$, and $D=1$. Measurements done in dimensionless units (A.U.).}
\label{Figure-6}
\end{figure}

Following the simulations of a single initial localized patch (Fig.~\ref{Figure-4}), 
we performed large-scale simulations, containing hundreds of randomly distributed patches, 
the objective of this, was to assess how fundamental is the self-replicating 
process in the emergence of the characteristic wavelength in the system. 
We proceeded as follows: (i) Considering a system of $256\times256$-points ($dx~=~2.0$), 
we built and initial condition consisting of a spatial Poisson point process with rate $0.001$, 
this generates on the order of 600 point randomly distributed in the two-dimensional plane. 
Each of the points is considered as the center of a localized patch of $3$-points radius, the generated state can be observed in Fig.~\ref{Figure-6}a. (ii) The random field is used as initial condition for simulating Eq.~(\ref{nonlocal}) for a parameter region inside the self-replication windows (see Fig.~\ref{Figure-5}, $L_c=~4.50$). (iii) We let the system evolve, and obtain a transient state, that can be seen in Fig.~\ref{Figure-5}c, although here we cannot observe new structures, because the evolution time was too short, it is enough time for the system to split the existing cluster into distinct structures.

When computing the Fourier transform of both the initial and the evolved state, we see that for the case of the initial state (Fig.~\ref{Figure-6}b) the Fourier transform (apart from the central peak) exhibits a characteristic wavelength of the order of the size of the localized structures ($\lambda~=~3$ A.U), due to the almost mono-dispersed sizes of the computer generated structures (almost, because some of the structures merged to generate bigger clusters), the same phenomenon was observable in the analysis of Zambia's landscape (Fig.~\ref{Figure-3}c). 

On the contrary, when analyzing the evolved state (Fig.~\ref{Figure-6}d), we see immediately the emergence of a characteristic wavelength much larger than the size of the structures ($\lambda~=~14$ A.U.), this wavelength emerges as a result of the self-organizing nature of the system, where clusters split, rapidly arriving  at a characteristic distribution.  

\subsection{Fragmentation: From homogeneous cover to patchy landscapes}
\begin{figure}[t!]
\centering
\includegraphics[width=12 cm]{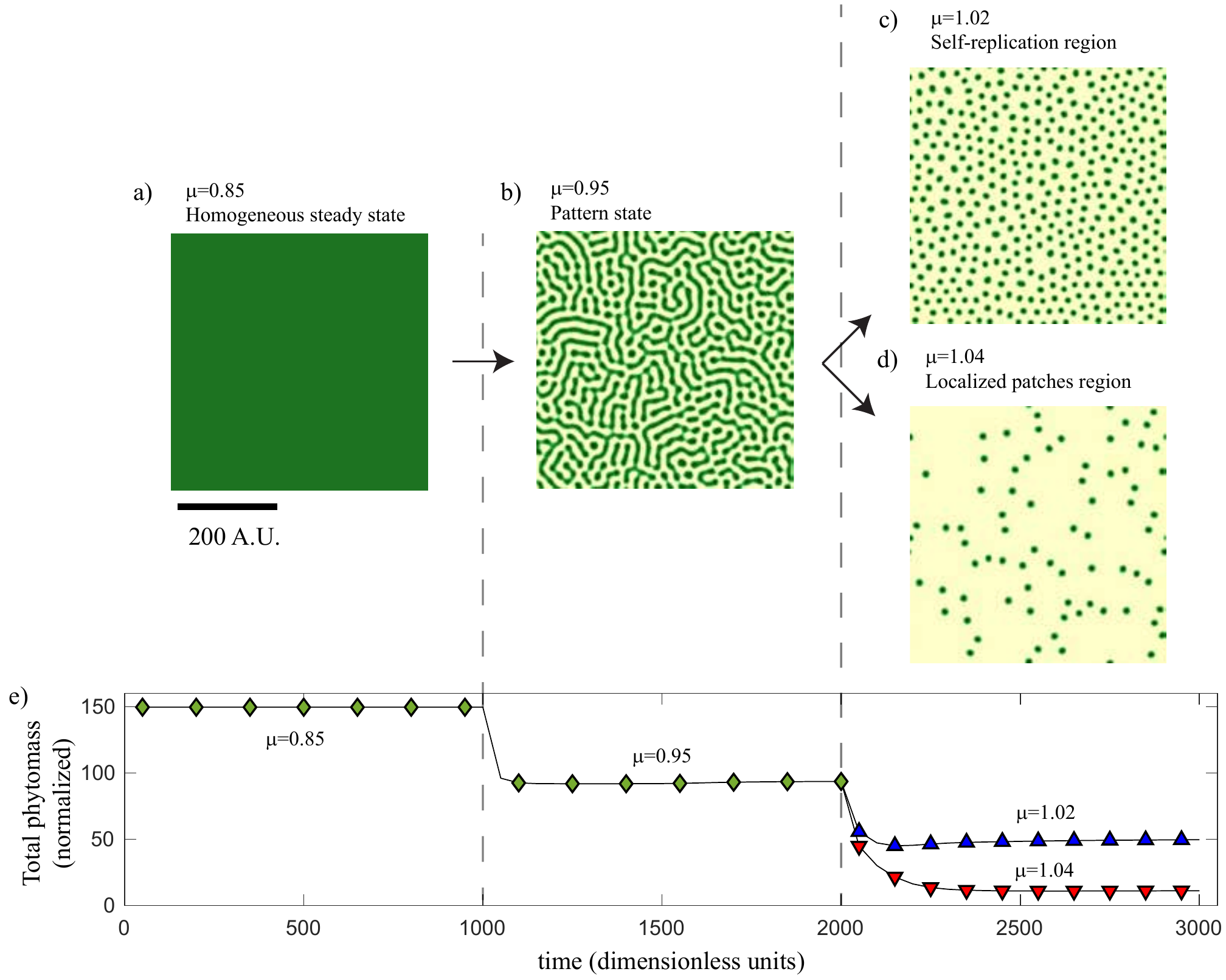}
\caption{(color online) {\bf Fragmentation: homogeneous cover decays as aridity increases}. a) $\mu=0.85$, initial condition, the system is in a completely vegetated state. b) $mu=0.95$ the homogeneous state decays to a labyrinthine pattern. c) $mu=1.02$, (self replicating region) labyrinthine pattern decays to a non-periodic patch pattern. d) $mu = 1.04$, (stable localized patches region) labyrinthine pattern decays to isolated patches without evident global order. e) Shows the temporal evolution of the total biomass (normalized by the biomass of a the single LP in Fig.~\ref{Evo1}), diamonds show the transition from $\mu=0.85$ to $\mu=0.95$, upward-triangles show the transition to $\mu=1.04$, while downward-triangles show the transition to $\mu=1.02$. Simulation parameters: $256\times256$-point grid ($dx=2.0$), $dt = 0.1$, $L_c=4.5$, $\chi_f=2$, $\chi_c=1$, and $D=1$.  Multiplicative noise with amplitude $0.01$ is used to facilitate the initial destabilization of the homogeneous state.}
\label{Evo2}
\end{figure}

We have show that through self-replication a single or multiple localized patches may increase the total biomass of an system leading ultimately to  a hexagonal patterned state. However this is not the only route by which a system can reach such patterned state, a second route is presented in Fig.~\ref{Evo2}, where an homogeneous vegetated state ($\mu=0.85$) is destabilized by an increase in aridity towards a labyrinthine pattern ($\mu=0.95$). Further increasing aridity drives the system towards two possible final states, the first possibility is a reduction in aridity that takes the system to the region where localized patches are unstable and suffer from self-replication ($\mu=1.02$), here, the labyrinthine pattern breaks down into multiple localized patches that cover the system by generating a patterned state. Note that the bare state become stable when $\mu>1$. On the other hand, when aridity is shifted from $\mu=0.95$ to $\mu=1.04$, where localized patches are stable, then the system reaches a different state, consisting on a low number of isolated patches distributed in a disordered manner.

It is important to mention that a discontinuous transition from the stable homogeneous cover region ($\mu=0.85$) directly to any of the non-periodic patch regions ($\mu=1.02$ or $1.04$) generates an overall decay of the homogeneous state to the bare state. Thus, an intermediate level of aridity is necessary for allowing the destabilization of the homogeneous cover into a non-periodic patch pattern.

\subsection{Remote-sensing observations}

\label{Sec-Remote}

We have studied the spatial distribution of localized patches in eight distinct regions, four located in the Fombeni region, Mozambique, and other four located in the Mufumbwe District, North-Western Province of Zambia. The vegetation in Zambia is dominated be a medium size tree {\it Brachystegia spiciformis}. The vegetation patterns observed in these regions are not periodic,  and are composed by vegetation patches, possibly composed by groups of trees forming compact clusters. Each patch can cover an area of up to thousands square-meters, with an effective radius of tens of meters, as can be observed in Figs.~\ref{Figure-1}a and \ref{Figure-1}f (and SI~Figs.~\ref{Fig-StatMoz} and \ref{Fig-StatZam}).

\begin{figure}[ht!]
\centering
\includegraphics[width=12 cm]{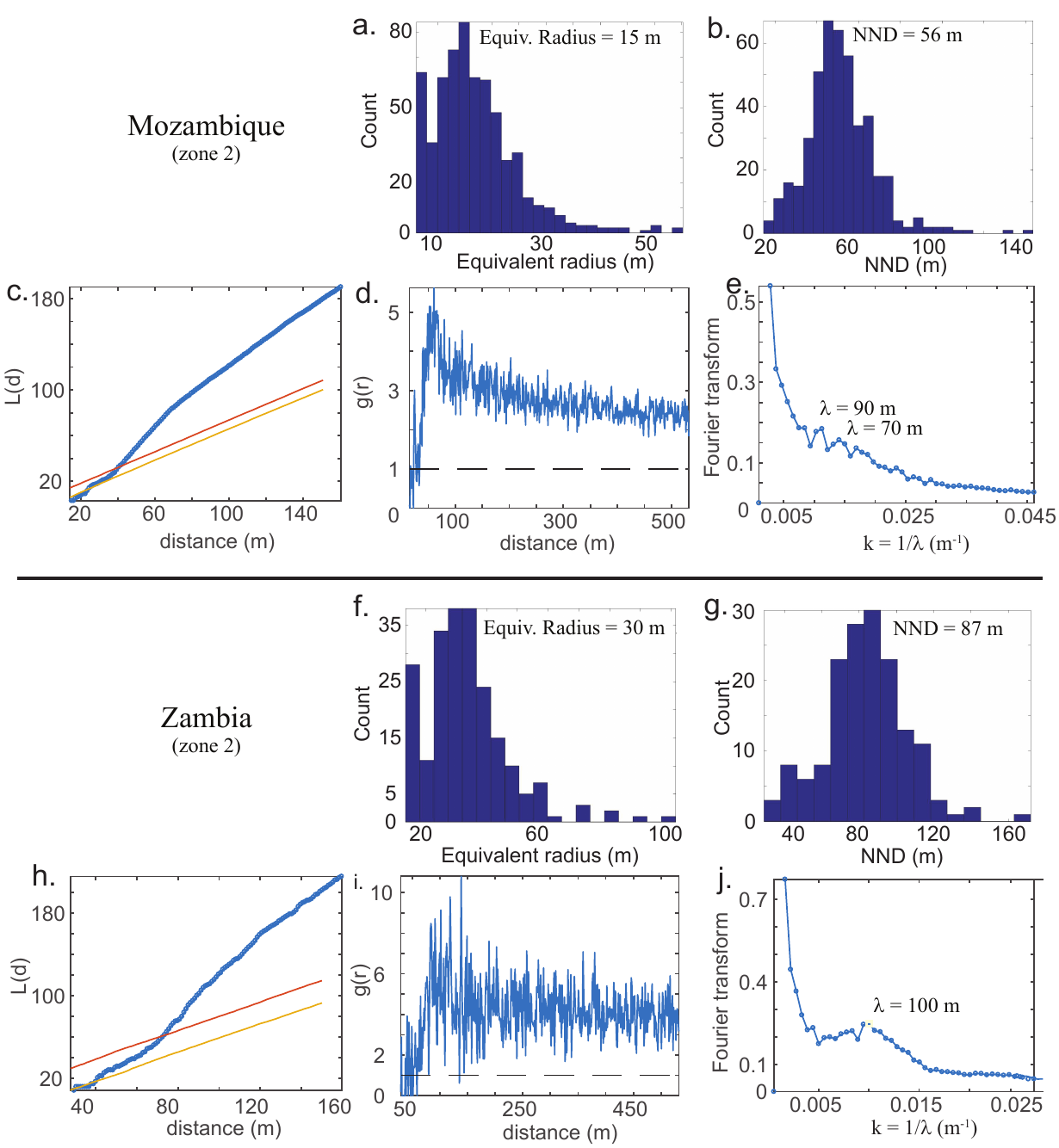}
\caption{(color online) {\bf Measurements performed on the detected patches} for {\it zones 2} of Mozambique (top) and Zambia (bottom). (a) \& (f) show the histograms for the equivalent radii of the structures and (b) and (g) show the histogram of the nearest-neighbor distances between them.  the modified K-function $L(r)$ (lines show the 95\% confidence interval) shown in (c) and (h), while (d) and (i) show the radial distribution function $g(r)$. Finally the circular average of the two-dimensional Fourier transform, i.e, the spectra are shown in (e) and (j).}
\label{Figure-2}
\end{figure}

The maximal values of the radius of the structures extends over hundreds of meters, 
however, the typical value (mode) varies from region to region, ranging from 7 meters (zones 3, and zone 4 in Zambia) 
up to 36 meters (Zambia zone 4), most of the regions show a marked characteristic size of the structures (see Fig.~\ref{Figure-2}a and \ref{Figure-2}f, and SI Figs.~\ref{Fig-StatMoz} and \ref{Fig-StatZam}) and nearest-neighbor distance (see Fig.~\ref{Figure-2}b and \ref{Figure-2}g), 
indicating a preferred spatial patch distribution.

To study the spatial  organization of the patterns, we perform two measurements, 
the first is a modified Ripley's function L, 
which measures the level of clustering of a spatial point process (see formula  \ref{Eq-L}), 
and the second one is the radial distribution function (\ref{Eq-g}), used to measure the correlation 
in a point process. Both observables are closely related \cite{Couteron_Kokou_97}, yet together they allow for further 
insight into the spatial organization. All regions Zambia and zones 1 and 4 in Mozambique exhibit scarcity of neighbors at close distance 
(observations below the 95\% confidence interval) and then transition towards clustering is observed 
for larger distances (Fig.~\ref{Figure-2}c and \ref{Figure-2}h). Similarly, when analyzing the correlation of the point 
distribution through the radial distribution function, we observe a positive correlation ($g(r)>1$) 
regime in all the zone, in particular, regions with high clustering exhibits long range positive correlation, 
while zones 1 and 3 in Mozambique, which show no clustering exhibit a positive correlation only for 
a small distance (Fig.~\ref{Figure-2}d and \ref{Figure-2}i).
Because of the size of the regions considered (see SI Fig.~\ref{Fig-Zones}) where the terrain is not 
homogeneous, and there is a qualitative change in the structures even within a single zone, this prevents the definition of a characteristic wavelength through a spatial Fourier transform on every region. Positive examples are 
zones 2 in Mozambique and Zambia, where an incipient characteristic wavelength can be defined (Fig.~\ref{Figure-2}e and \ref{Figure-2}j).

\begin{figure}[ht!]
\centering
\includegraphics[width=8 cm]{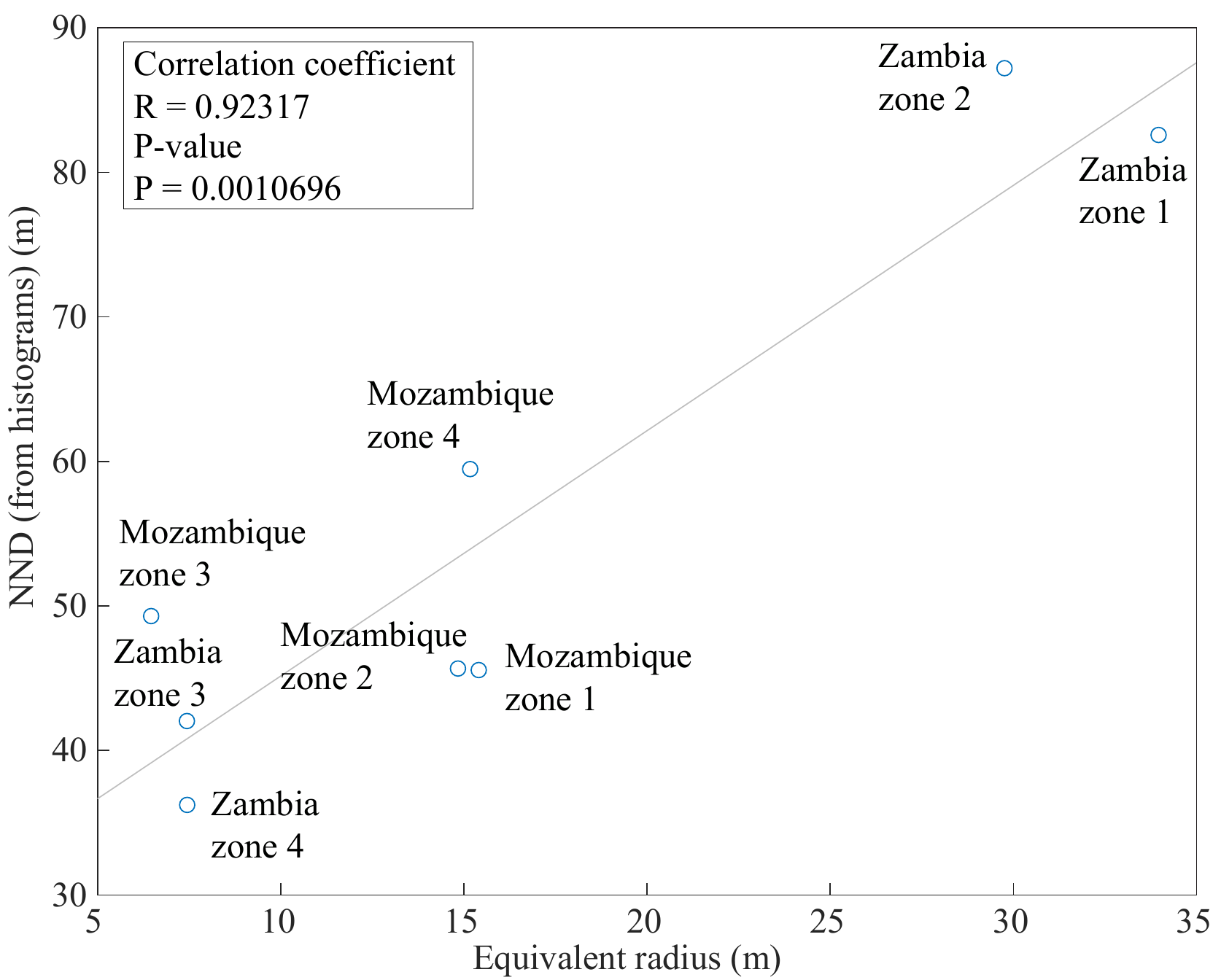}
\caption{(color online) {\bf Correlation between the mean equivalent radius of the structures and 
the nearest-neighbor distance} for each of the eight regions analyzed, there is a positive linear 
correlation between these two quantities ($P<0.05$), which allow us to discard the null hypotheses.}
\label{Figure-3}
\end{figure}

To study if there is a relationship between the size of the patches and the distance between them, we make use of the results from the analysis of radius and nearest-neighbor distance (NND) between structures (see SI Figs.~\ref{Fig-StatMoz} and \ref{Fig-StatZam}) 
and study the correlation between them through a linear correlation analysis (see Fig.~\ref{Figure-3}).
We observe  a strong correlation between the two quantities, indicating that there might be 
a underlying mechanism controlling both the structures size and the distance between them.

\section{Discussion}
Numerical observation of the non-local model studied suggest that self-replication may be an important path for the propagation of vegetation, which allows for the propagation of the vegetation even in arid conditions where the homogeneous cover is not viable due to the lack of resource (nutrients and/or water). Moreover, it has been shown that a transition from a homogeneous vegetation cover to a patch patterned state may only occur if the transition occurs towards the self-replicating region. However, in real-ecosystems processes occur at a slow time scale and there is a high number of uncontrolled parameters which prevent the assessment of the validity of the numerical observations. We however illustrate the plausibility of the process by referring to observations at a single point in time of patchy vegetation in Africa.  

The properties of the landscapes observed in Mozambique and Zambia show some indications that support self-organization. These systems exhibit a characteristic distances between patches, and a characteristic patch size, together with clustering and long range correlations in the spatial distribution. This may indicate that 
patches are interacting with each other resulting in the spatial distributions observed. From our modeling perspective, localized patches which are too 
close to each other (high density) will compete for resources 
ending up with the disappearance of the one placed in the less favourable location. If the patches are too far apart (low density), the extra space will allow the growth of the patches resulting in self-replication and the formation of new patches that will fill the previously unoccupied region. We must here make clear that we cannot corroborate the existence of these dynamical processes from the data analysis performed. The remote sensing data provides no temporal information and limited spacial resolution. Further on-site investigations are necessary to comprehend plant-plant interaction in specific contexts, and to assess if there are local factors that contribute to the  self-organization of the landscape.

Despite all of this, we have observed features  that support the patch self-replication hypothesis, for example, the observation of rings, which is a second route to self-replication (see Fig.~\ref{Figure-5}). Rings are usually transient state, leading to division into two or more patches \cite{bordeu2016self,meron2004vegetation}. Examples of this can be observed in different regions of Zambia (for example, [$14^\circ40'10.36''$S, $25^\circ 49'37.34''$E], and the surrounding areas). It is harder to observe the formation of rings in Mozambique, and we believe this relates to the facts that the structures in this location are smaller, preventing the ring instability to take place. 
Moreover the direct correlation between the size of the patches and the nearest neighbor distance, is also rendered by the model, which indicates that there is a dominant  wavelength in the system that controls both the size of the structures and distance between them. This wavelength depend on the aridity parameter (see Appendix~\ref{An-1}).

\section{Conclusions and perspectives}
By making use of a generic interaction redistribution model, we showed that self replication of localized patches is generically present even in the most general model involving non-local competition and facilitating interactions. We have shown, through this model, that a localized structure may undergo multiple self-replications to finally cover the whole space available with a regular hexagonal pattern, and by doing so, let the total biomass in the system increase in time. Moreover, the self-replication regime is may be an important stage in the decay of homogeneous covers of vegetation to hexagonal pattern states. It may be seen as an alternative to a more abrupt decay in biomass without this, the decay in the biomass is more abrupt resulting in a low number of sparse patches. We hypothesize that these processes are occurring in the observed landscapes of Mozambique and Zambia.

Through the analysis of satellite images, we have given evidence to support the self-organization hypothesis. Despite the limited amount of data, which prevents being assertive about the underlying processes, we believe that self-replication could be considered as a possible mechanism for the regeneration dynamics of these landscapes. This phenomenon may occur for a moderate level of the aridity. Moreover, when considering a homogeneous cover, a gradual increasing of the level of the aridity results in a decay of the total biomass. The degradation of the ecosystems though fragmentation leads to either a periodic vegetation patterns or a random distribution of localized patches of vegetation. However, a periodic pattern is only obtained if the system relaxes in the region where self-replication exists, and patches are unstable.

As mentioned in the Discussion section, more data, specially field measurements, long time-lapse imaging and controlled 
experiments are necessary to confirm any claim on the nature of the observed patterns. However, 
 measurements done here are a first step to understand the distribution of the 
observed patterns, and illustrate that self-organization is a strong candidate to explain them. Despite that 
there is no direct observation of self-replication,  the current state of some patches drive 
us to think that self-replication might be an ongoing process, for example in the "dividing" 
patches showed in Fig.~\ref{Figure-1}. 

The knowledge of the below-ground structures of the type of vegetation considered in this contribution is rather limited. From the current literature  only rooting depth  data and a quantitative index of the vertical distribution of roots  are available. which are irrelevant for the present study. Moreover, the full lateral roots extension has not yet been measured. For this reasons it was beyond the scope of the present contribution to perform a quantitative comparison between theoretical results and field observations. On the other hand, the model presented here has been previously confronted  quantitatively to experimental measurement for two type of plants (Festuka orthophylla \cite{Couteron20140102} and Combretum micranthum \cite{JEC:JEC1126,ECY:ECY20088961521,Lefever2009194}. We believe that the same measurement could be realized for the type of vegetation of Zambia and Mozambique. 

\section{Appendix}
\subsection{Stability analysis of the integro-differential model \label{An-1}}
\begin{figure}[ht!]
\centering
\includegraphics[width=12 cm]{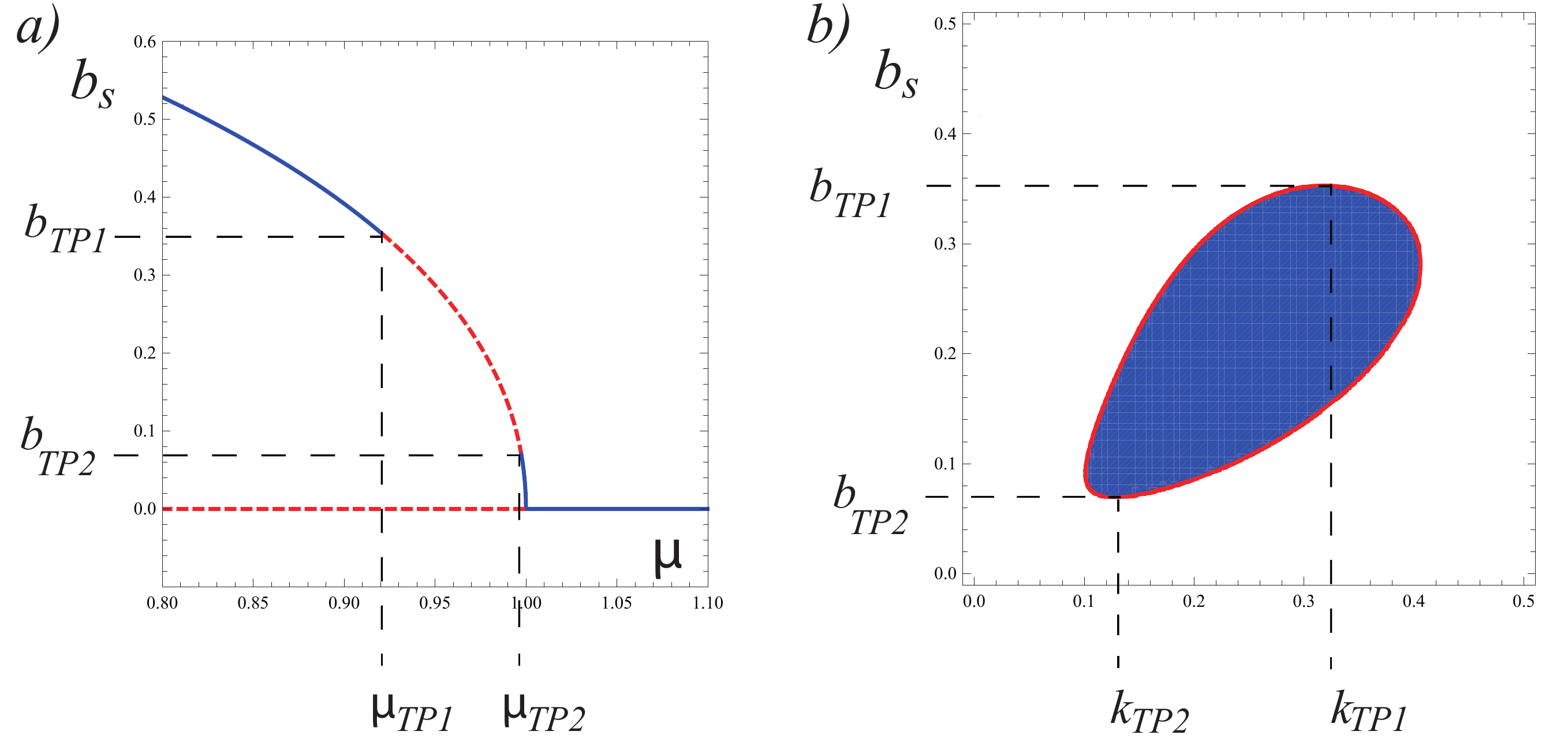}
\caption{(color online) {\bf Stability diagrams.} (a) Uniform stationary distributions of the biomass density $b_s$, 
and their stability with respect to homogeneous and inhomogeneous perturbations 
are plotted as a function of the aridity parameter $\mu$. Stable states 
are indicated by solid line, and unstable ones are represented by dotted lines. (b) 
The marginal stability curve in the ($b_s$-$k$) plane. The domain of instability for  
is represented by a blue shaded area delimited by red solid line. Parameter are 
$\chi_f = 2$, $\chi_c = 1$, $L_c = 4$, and $D=1$.  }
\label{Figure-Linear}
\end{figure}

\begin{figure}[ht!]
\centering
\includegraphics[width=10 cm]{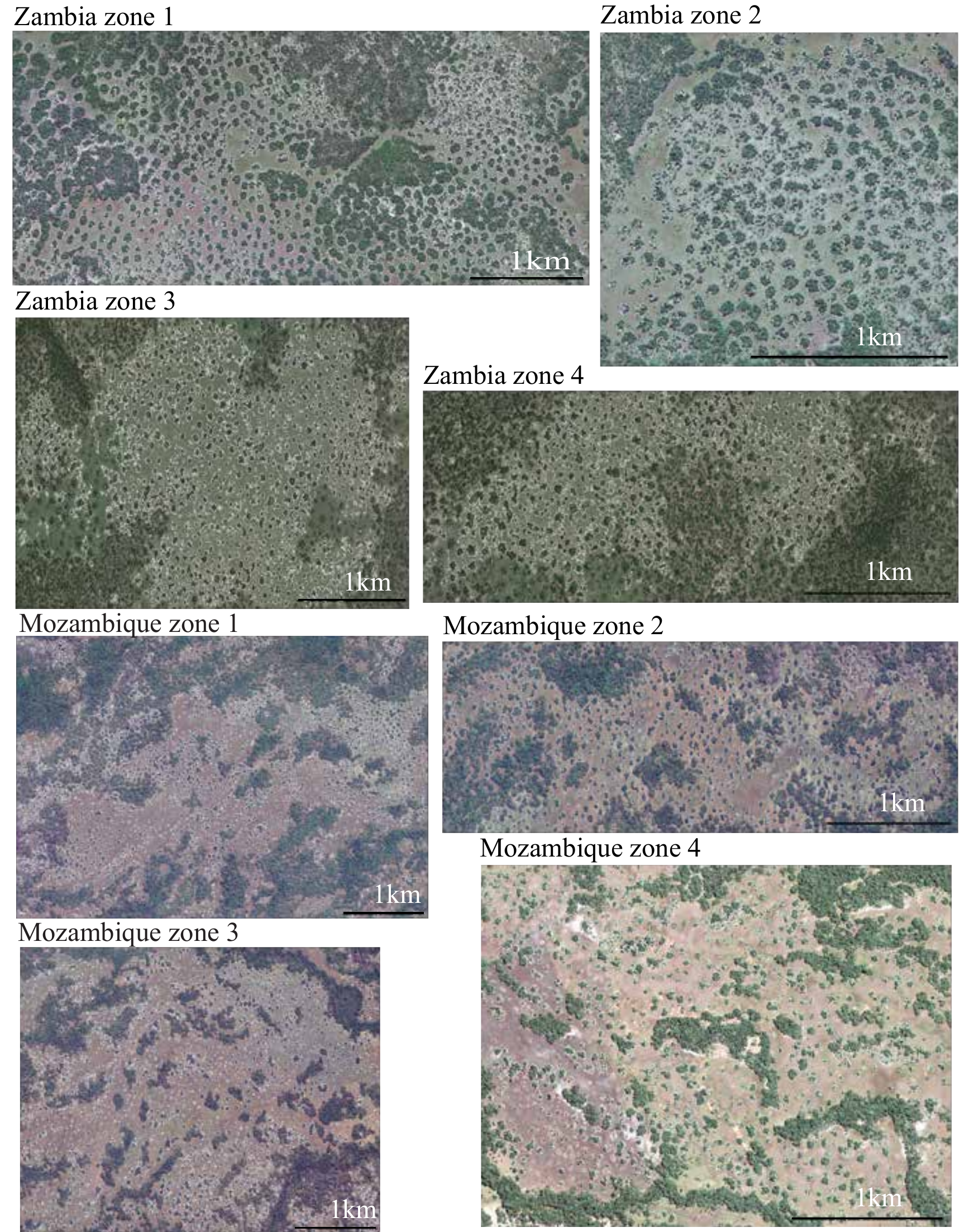}
\caption{{\bf Zones considered for the analysis}, Cordinates: 
Zambia zone 1 [$ 13^{\circ} 47'23.99''$S, $ 25^\circ17'11.18''$E],
Zambia zone 2 [$ 13^{\circ} 51'17.88''$S, $ 25^\circ22'33.86''$E],
Zambia zone 3 [$ 14^{\circ} 39'16.49''$S, $ 25^\circ49'54.86''$E],
Zambia zone 4 [$ 14^{\circ} 40'21.72''$S, $ 25^\circ49'38.56''$E],
Mozambique zone 1 [$ 18^{\circ} 41'19.32''$S, $ 35^\circ30'37.62''$E],
Mozambique zone 2 [$ 18^{\circ} 40'23.96''$S, $ 35^\circ33'32.58''$E],
Mozambique zone 3 [$ 18^{\circ} 41'55.52''$S, $ 35^\circ38'14.45''$E],
Mozambique zone 4 [$ 18^{\circ} 49'48.31''$S, $ 35^\circ36'22.37''$E]. The white boundaries enclosing each structure corresponds to results of the automatic detection of objects.}
\label{Fig-Zones}
\end{figure}

\begin{figure}[htbp!]
\centering
\includegraphics[width=12 cm]{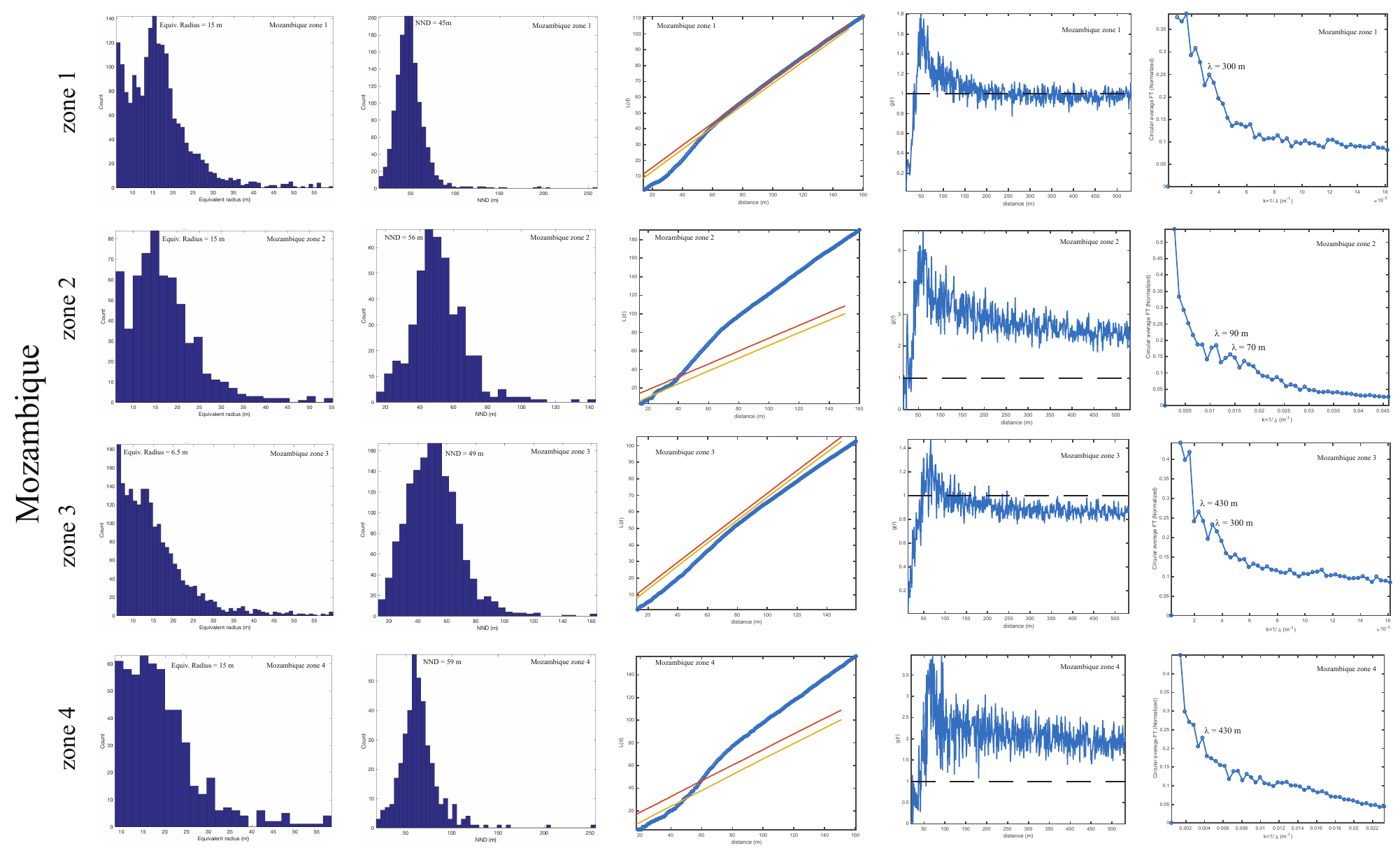}
\caption{{\bf Analysis of the zones in Mozambique}, each row corresponds to the measurements on a different zone (1 to 4, from top to bottom). From left to right: Histogram of equivalent radius, histogram of nearest-neighbor distance, L-function, radial distribution function, and circular average of the spatial Fourier transform, respectively.}
\label{Fig-StatMoz}
\end{figure}

\begin{figure}[htbp!]
\centering
\includegraphics[width=12 cm]{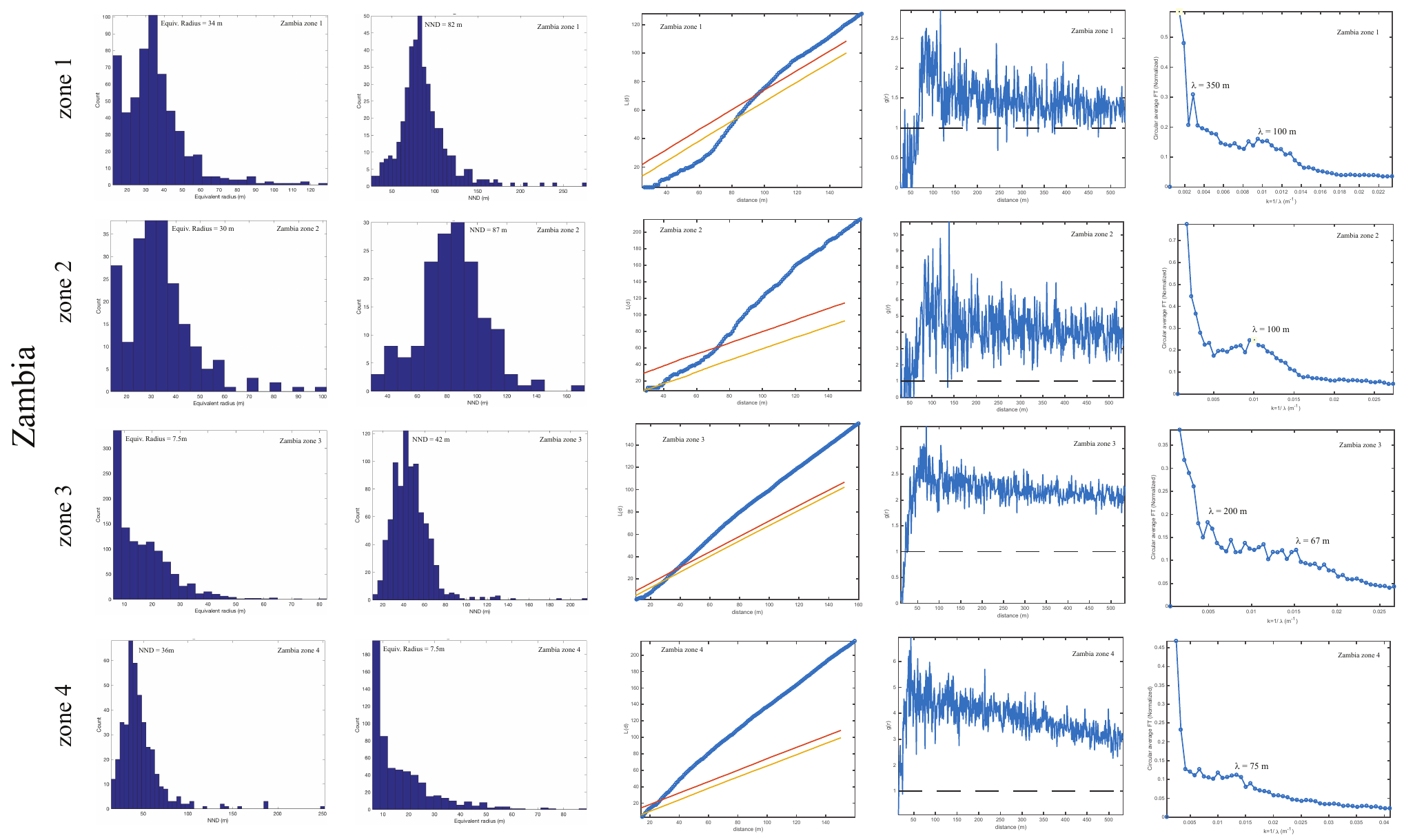}
\caption{{\bf Analysis of the zones in Zambia}, each row corresponds to the measurements on a different zone (1 to 4, from top to bottom). From left to right: Histogram of equivalent radius, histogram of nearest-neighbor distance, L-function, radial distribution function, and circular average of the spatial Fourier transform, respectively.}
\label{Fig-StatZam}
\end{figure}

The stationary homogeneous states of Eq. (\ref{nonlocal}), describing uniform vegetation covers, 
are  $b_0 = 0$, and $b_s$, given by the solution of
\begin{equation}\label{HSS} \mu=(1-b_s)\exp(\Lambda b_s),
\end{equation}  
where the feedback difference $\Lambda\equiv\chi_{f}-\chi_{c}$ measures the community 
cooperativity. The first, $b_0 = 0$, represents a territory totally devoid of vegetation, i.e., a 
bare state. The second, $b_s$ corresponds to uniform plant distributions that can be either monostable 
if $\Lambda<1$ or bistable when $\Lambda>1$. In the monostable case, $\Lambda<1$, the biomass 
density decreases monotonously with the aridity parameter $\mu$ and vanishes 
at $\mu=1$ (cf. Fig.~\ref{Figure-Linear}). It exist only if $\mu<1$. In the bistable case, 
$\Lambda>1$, the uniform plant distributions 
extends up to the turning point sometimes also referred to as tipping point (saddle-node bifurcation point). 
The coordinates of this point are $b_l=(\Lambda-1)/\Lambda$ and $\mu_l=\exp{(\Lambda -1)}/\Lambda$.

To study the  linear stability analysis of the stationary homogeneous states, $b_s$, 
we introduce a small amplitude deviations from $b_s$ of the form
\begin{equation} \label{pertur}
b({\bf{r}},t) = b_s + \varepsilon \exp\left(\lambda(k) t + i {\bf{k}}\cdot{\bf{r}}\right),
\end{equation}
where $\varepsilon$ is a small parameter and the deviation from the $b_s$ is  expressed in 
terms of Fourier modes  $\exp\left(\lambda(k) t + i {\bf{k}}\cdot{\bf{r}}\right)$ in the space of wavevector ${\bf k}$. 
By replacing Eq.~(\ref{pertur}) and by linearizing 
with respect to $\varepsilon$, the dispersion relation obeyed by the eigenvalues $\lambda(k)$ reads 
\begin{equation} \label{spectrum2}
\lambda(k) = \left[\chi_f(1-b_s) - 1 - \frac{\chi_c (1-b_s) }{\left(1 + (kL_c)^2\right)^{3/2}}\right]b_s e^{\chi_f b_s} -Dk^2.
\end{equation}
The critical points associated with the Turing-Prigogine instability are given by, 
\begin{eqnarray}
\left.\frac{\partial \lambda}{\partial k}\right|_{b_s = b_c,~k=k_c} &=& 0, \label{C1}\\
\lambda(k_c) &=& 0. \label{C2}
\end{eqnarray}
These two conditions provide the thresholds $b_c=b_{TP}$ and the most unstable wavenumber $k_c= k_{TP}$ 
associated with the Turing-Prigogine instability. The critical wavenumber at the onset of the instability is given explicitly by

\begin{equation} \label{kc2d}
k_{TP}^2 = \frac{1}{L_c^2}\biggl[\left(\frac{\chi_c b_{TP}(1-b_{TP}) e^{\chi_f b_{TP}}L_c^2}{2D}\right)^{\frac{2}{5}} - 1\biggl].
\end{equation}
The wavelength is $2\pi/k_{TP}$.  The above equation has two solutions, 
which we denote by $(k_{TP1},b_{TP1})$
and $(k_{TP2},b_{TP2})$.
The thresholds associated with the Turing-Prigogine instability are solutions of
\begin{equation} \label{bc2d}
\left [\chi_f(1-b_{TP}) - 1\right]e^{\chi_f b_{TP}} = 3\left(\frac{D}{L_c^2} \right)^{\frac{3}{5}}
\left[\frac{\chi_c b_{TP}(1-b_{TP}) e^{\chi_f b_{TP}}}{2} \right]^{\frac{2}{5}}  - \left(\frac{D}{L_c^2} \right).
\end{equation}
The results of the linear stability analysis are summarized in  
Fig. \ref{Figure-Linear}. As the aridity level  $\mu$ increases, the vegetation evolves toward 
extinction. For $0<\Lambda \leq 1$, the decrease of the phytomass density is monotonous. 
The bare state density is reached at the switching point $\mu=1$, where the plant distribution state is destabilized through the Turing-Prigogine instability \cite{turing1952chemical,prigogine1968symmetry}. In the range  $b_{TP1}<b<b_{TP2}$, the uniform phytomass state is unstable. 
The corresponding levels of the aridity $\mu_{TP}$ are obtained from the homogeneous 
steady states Eq.~(\ref{HSS})  (see Fig.~\ref{Figure-Linear}).

When increasing the stress or aridity, the wavelength of the corresponding pattern increases, 
and simultaneously the morphology of the vegetation pattern changes, this is related to the 
observations made in Mozambique and Zambia, where an increase in 
the system wavelength is directly correlated with an increase in the size of the vegetation patches. The generic sequence of spatially periodic states obtained for increasing aridity is: spots 
of lower density forming an hexagonal lattice, alternating stripes of higher and 
lower density, and spots of higher density forming an hexagonal lattice \cite{LT,QUA:QUA10878}. 
This behavior has been also found for other mathematical models  that include water transport by 
underground diffusion and/or above ground run-off \cite{Meron2001,Rietkerk2008169,Sherratt20138,Gowda}.

\section*{Acknowledgements}
We acknowledge fruitful discussions with Ren\'e Lefever and Pierre Couteron. We also appreciate the invitation from Pierre Couteron to submit this contribution to Ecological Indicators.
M.T. received support from the Fonds National de la Recherche Scientifique (Belgium).
I.B. was supported by CONICYT, Beca de Doctorado en el Extranjero No. 72160465.
 M.G.C. thanks the financial support of FONDECYT project 1150507.  M.G.C. and M.T. acknowledge
the support of CONICYT project REDES150046.  This research was also
supported by Wallonie-Bruxelles International (WBI).

\pagebreak
\section*{Supplementary information}
\subsection*{Areas of study}

\begin{table}\label{Table-PP}
\centering
\begin{tabular}{|c | c| c| c|}
\hline
Location & \begin{tabular}{@{}c@{}}Rainfall \\ mm/year\end{tabular} & \begin{tabular}{@{}c@{}}Mean temp. \\ 1901-2006\end{tabular} & \begin{tabular}{@{}c@{}}Mean temp. \\ 2006-2016\end{tabular}  \\
\hline
Zambia zone 1 & 903 & $21.99^\circ$ & $22.72^\circ$ \\
Zambia zone 2 & 903 &  $21.99^\circ$ & $22.72^\circ$ \\
Zambia zone 3 & 903 &  $21.99^\circ$ & $22.72^\circ$ \\
Zambia zone 4 & 903 &  $21.99^\circ$ & $22.72^\circ$ \\
Mozambique zone 1 & 1348 & 24.89$^\circ$ & $25.49^\circ$ \\
Mozambique zone 2 & 1348 & 24.89$^\circ$ & $25.49^\circ$ \\
Mozambique zone 3 & 1348 & 24.89$^\circ$ & $25.49^\circ$ \\
Mozambique zone 4 & 1348 & 24.89$^\circ$ & $25.49^\circ$ \\
\hline
\end{tabular} 
\caption{{\bf Information for analyzed regions}. Mean annual rainfall, and mean temperature for the periods 1901-2006 and 2006-2016 (Obtained from CRU TS3~\cite{CRU}) are shown.}
\end{table}

\subsection*{Measured properties}

\subsection*{Algorithm}
\label{Code}
Here we present the algorithm written for the detection of vegetation patches in the eight regions selected, the parameters used for each region analyzed are indicated within the code.
\begin{lstlisting}[frame=single,basicstyle=\tiny,numbers=none,showspaces=false,showstringspaces=false]
%% Ecological Indicators
%% Large scale patchy ecosystems may increase total biomass through a self-replication process.
%% Mustapha Tlidi, Ignacio Bordeu, Marcel G. Clerc, Danel Escaff.

% This algorithm detects and extracts the properties and plots the objects of
% interest, for that it uses two functions: Detection and plots. Both are
% described in detail below.

% To run, save the image files and the algorithm in the same folder, open
% the algorithm in matlab and type spotDetection() on the command window.
% [maxNND,maxReq] = spotDetection()
%-----------------------------------------------------%
%-----------------------------------------------------%
function spotDetection()
% This function of just the wrapper, it loads the images and calls 
% Detection to obtain the properties of the objects.
Z1 = imread('zone1_Zambia.jpg'); 
Z2 = imread('zone2_Zambia.jpg');
Z3 = imread('zone3_Zambia.jpg');
Z4 = imread('zone4_Zambia.jpg');
M1 = imread('zone1_Mozambique.jpg');
M2 = imread('zone2_Mozambique.jpg');
M3 = imread('zone3_Mozambique.jpg');
M4 = imread('zone4_Mozambique.jpg');
%-----------------------------------------------------%
% The parameters for the detection are chosen in order to 
% maximize the detection of patches 
[statsZ1,imSZ1,BZ1,LZ1] = Detection(Z1,10,'red',400,50000); 
[statsZ2,imSZ2,BZ2,LZ2] = Detection(Z2,10,'red',500,50000);
[statsZ3,imSZ3,BZ3,LZ3] = Detection(Z3,5,'green',100,50000);
[statsZ4,imSZ4,BZ4,LZ4] = Detection(Z4,5,'green',100,50000);
[statsM1,imSM1,BM1,LM1] = Detection(M1,100,'red',100,10000);
[statsM2,imSM2,BM2,LM2] = Detection(M2,50,'red',100,10000);
[statsM3,imSM3,BM3,LM3] = Detection(M3,50,'red',100,10000);
[statsM4,imSM4,BM4,LM4] = Detection(M4,200,'red',200,10000);
%-----------------------------------------------------%
% The properties obtaines in 'stats' must be scaled according to dx, the
% meter-to-pixel ratio:
dx = 300/282; % meters/pixel extracted from the images
%-----------------------------------------------------%
% Finnaly, we plot all the merge of the original image and the detected 
% objects 
plots(Z1,BZ1,name{1})
plots(Z2,BZ2,name{2})
plots(Z3,BZ3,name{3})
plots(Z4,BZ4,name{4})
plots(M1,BM1,name{5})
plots(M2,BM2,name{6})
plots(M3,BM3,name{7})
plots(M4,BM4,name{8})
end

function [statsSpots,imSpots,BSpots,LSpots] = Detection(imRGB,color_thresh,use_channel,min_size,max_size)
% This funtion detects the objects of interest in the RBG input image im
% im : Input RGB image
% color_thresh : only pixels with a value lower than this are considered
% use_channel : channel used for the thresholding 'red' or 'green'
% min_size : minimum size in pixels of structures to consider
% max_size : maximum size in pixels of structures to consider
%-----------------------------------------------------%
%-----------------------------------------------------%
% We make use of the fact that the terrain is of an orange tone, 
% contrasting to the green color of the vegetation.
% We adjust the intensities of the red and green channels of the image to 
% increase the contrast between the backround and the vegetation. Here, we 
% are effectivelly eliminating (almost) any contribution of the red channel 
% to the vegetation and green contribution to the background.
imadjusted = imadjust(imRGB,[.4 .4 0;.6 .6 1],[]);
%-----------------------------------------------------%
% We extract the green (2nd channel) and the adjusted red channel (1st ch).
redCh = imadjusted(:,:,1);
greenCh = imadjusted(:,:,2);
%-----------------------------------------------------%
% and construct a binary image where the value of 1 correspond to the
% pixels where the level of red (or green) is below color_thresh 
% (the scale is [0 255]), this threshold is defined by inspection of the 
% real image.
switch use_channel
    case 'red'
        imbw = redCh<color_thresh;
    case 'green'
        imbw = greenCh<color_thresh;
end
%-----------------------------------------------------%
% We proceed by deleting objects touching the edges of the binary image
% imbw, and also filling any holes in the detected objects
imclear = imclearborder(imbw);
imfull = imfill(imclear,'holes');
% and also deleteng very small structures, with area below 100 pixels and
% also deleting objects with an area above 10000 pixels, these threshold
% are also set by inspection
imSpots = xor(bwareaopen(imfull,min_size),bwareaopen(imfull,max_size));
%-----------------------------------------------------%
% we extract the Boundaries, and Labels of the structures
[BSpots,LSpots] = bwboundaries(imSpots,'noholes');
% and determine their individual properties that we use for the analysis of
% area, NND, equivalent diameter, etc.
statsSpots = regionprops(LSpots,'Centroid','EquivDiameter','Area',...
    'MajorAxisLength','MinorAxisLength','Perimeter','Eccentricity');

%-----------------------------------------------------%
%-----------------------------------------------------%
end

function plots(Im,B)
% This function plots the RGB image Im, together with the boundaries B of
% the objects detected
%-----------------------------------------------------%
    figure;
    image(Im)
    hold on
    for i=1:numel(B)
        b = B{i};
        plot(b(:,2),b(:,1),'w');
    end
end
\end{lstlisting}

\end{document}